\documentclass[manuscript]{aastex}
\usepackage{natbib}
\usepackage{booktabs}
\usepackage{threeparttable}
\usepackage{amsmath,bm}
\begin{document}

\title{Effects of alpha$-$proton differential flow on proton temperature anisotropy instabilities in the solar wind: \emph{Wind} observations}

\author{G. Q. Zhao\altaffilmark{1,2}, H. Li\altaffilmark{3,4}, H. Q. Feng\altaffilmark{1}, D. J. Wu\altaffilmark{5}, H. B. Li\altaffilmark{1}, A. Zhao \altaffilmark{1}}

\affil{$^1$Institute of Space Physics, Luoyang Normal University, Luoyang, China}
\affil{$^2$Henan Key Laboratory of Electromagnetic Transformation and Detection, Luoyang, China}
\affil{$^3$State Key Laboratory of Space Weather, National Space Science Center, CAS, Beijing, China}
\affil{$^4$University of Chinese Academy of Sciences, Beijing, China}
\affil{$^5$Purple Mountain Observatory, CAS, Nanjing, China}

\begin{abstract}
Plasma kinetic waves and alpha$-$proton differential flow are two important subjects on the topic of evolution of the solar wind. Based on the \emph{Wind} data during 2005$-$2015, this paper reports that the occurrence of electromagnetic cyclotron waves (ECWs) near the proton cyclotron frequency significantly depends on the direction of alpha$-$proton differential flow ${\bm V}_d$. As ${\bm V}_d$ rotates from anti-Sunward direction to Sunward direction, the occurrence rate of ECWs as well as the percentage of left-handed (LH) polarized ECWs decreases considerably. In particular, it is shown that the dominant polarization changes from LH polarization to right-handed polarization during the rotation. The investigation on proton and alpha particle parameters ordered by the direction of ${\bm V}_d$ further illustrates that large kinetic energies of alpha$-$proton differential flow correspond to high occurrence rates of ECWs. These results are well consistent with theoretical predictions for effects of alpha$-$proton differential flow on proton temperature anisotropy instabilities.
\end{abstract}

\keywords{Sun: solar wind -- waves -- instabilities -- interplanetary medium}

\section{Introduction}
The investigation of kinetic waves as well as kinetic instabilities is believed to be inherently important with topic on energizing particles and/or modifying their velocity distributions in a collisionless plasma \citep[e.g.,][]{hol75p63,mar06p01}. Especially, electromagnetic cyclotron waves (ECWs) near the proton cyclotron frequency are perhaps of particular interest; theoretical studies show that they can efficiently contribute their energy to particles or absorb energy from the particles through wave$-$particle resonant interactions \citep{mar82p30,huy99p45,hej15p31,hej18p48,woo18p49}. They have been extensively studied in various space environments, such as planetary magnetosphere \citep[e.g.,][]{rus07p23,rod10p07}, magnetosheath \citep[e.g.,][]{sch96p34,sou15p38}, and terrestrial foreshock regions \citep[e.g.,][]{smi85p29,won91p85}. In case of the solar wind, research on ECWs obtains a lot of interest in recent years \citep[e.g.,][]{jia09p05,jia10p15,jia14p23,boa15p10,gar16p30,xia18p64,xia18p08,liq19p55,zha19p75}. According to observations, in the case of parallel propagation and at scales close to the proton gyro-frequency, a noticeable result is that left-handed (LH) polarized ECWs are almost always the dominant waves in the solar wind \citep[e.g.,][]{boa15p10,zha18p15}. The polarization is described in the spacecraft frame and with respect to the direction of the background magnetic field throughout the paper except where noted. Theoretically, two kinetic instabilities driven by proton temperature anisotropies can contribute to the generation of the ECWs consisting of \textbf{LH} proton cyclotron waves and right-handed (RH) magnetosonic waves in the plasma frame \citep{gar76p41,gar93,kas02p39,mar04p02,hel06p01,omi14p42,Gar15p49,yoo17p04}. Proton cyclotron instability can be excited to produce cyclotron waves in a plasma with proton perpendicular temperature ($T_{\perp}$) larger than the parallel temperature ($T_\parallel$), while parallel firehose instability may arise to generate magnetosonic waves in a plasma with a converse temperature anisotropy ($T_{\perp} < T_\parallel$).

On the other hand, the phenomena of differential flow between alpha particles and protons in the solar wind, revealed in 1970s \citep[e.g.,][]{rob70p78,asb76p19,mar82p35}, have also attracted much attention in the context of plasma kinetic instabilities \citep[e.g.,][]{gar00p55,gar00p20,lix00p83,luq06p01,ver13p63}.
For effects of the differential flow on proton temperature anisotropy instabilities, it has been investigated by linear Vlasov$-$Maxwell theory \citep{pod11p41}, and by hybrid simulation \citep{hel06p07,mar18p53}. These studies demonstrated that (1) the presence of alpha$-$proton differential flow contributes to a larger growth rate of the proton cyclotron and parallel firehose instabilities; (2) it can break the symmetry of the unstable waves for their propagation directions so that proton cyclotron (parallel firehose) instability preferentially generate cyclotron (magnetosonic) waves propagating parallel (antiparallel) the direction of the differential flow vector ${\bm V}_d$. In particular, result (2) above was employed to explain the domination of LH polarization of ECWs with an assumption that ${\bm V}_d$ points outward from the Sun \citep{pod11p41,zha17p08,zha19p75}.

Note that, according to the theory \citep{pod11p41}, one can deduce that the dominant polarization should be RH polarization once ${\bm V}_d$ is observed actually to be toward the Sun. However, this deduction, to the best our knowledge, has not been examined by any in situ observation.
Moreover, the presence of ${\bm V}_d$ with a direction toward the Sun is possible, especially for the slow solar wind \citep[e.g.,][]{fuh18p84}. In this regard, some investigation on the occurrence of ECWs with various directions of ${\bm V}_d$ should be desirable.

In this paper, we report our finding that the occurrence of ECWs in the solar wind shows clear dependence on the radial angle of ${\bm V}_d$. In particular, LH ECWs can become secondary with a percentage much lower than that of RH ECWs when ${\bm V}_d$ is directed toward the Sun, which should provide crucial indication for the effect of alpha$-$proton differential flow. The data and analysis methods used in this paper are described in Section 2. The results are presented in Section 3. Section 4 is the discussion and conclusion.

\section{Data and analysis methods}
The data used in the present paper are based on the \emph{Wind} mission, which is a comprehensive solar wind laboratory in a halo orbit around the L1 Lagrange point. The magnetic field data are from the Magnetic Field Investigation (MFI) instrument sampled at a cadence of 0.092 s \citep{lep95p07}, and the plasma data are from the Solar Wind Experiment (SWE) instrument working at a cadence of 92 s \citep{ogi95p55}. The plasma data can give the ion (proton and alpha particle) bulk velocity and the perpendicular and parallel temperatures with respect to the background magnetic field; these ion data are produced via a nonlinear-least-squares bi-Maxwellian fit of ion spectrum from the Faraday cup \citep{kas06p05}. The differential flow vector is defined as ${\bm V}_d = {\bm V}_\alpha -{\bm V}_p$ in this paper, where ${\bm V}_\alpha$ and ${\bm V}_p$ are proton and alpha particle bulk velocities, respectively.

It is well acknowledged that the solar wind coming from different source regions on the solar surface often has different physical situations \citep[e.g.,][]{xuf15p70}. In order to reduce the possible combined effect of solar winds with different origins, a categorization of the solar wind should be appropriate. A traditional categorization approach is frequently based on the solar wind speed, but studies show that the speed is not necessarily a good parameter for characterization of the solar wind \citep{mar81p99,sta15p00,ami19p65,sta19p06}. Instead of the traditional approach, the categorization in this paper is conducted by an eight-dimensional scheme for 4-type solar wind categorization based on the machine learning technique with $k$-Nearest Neighbor classifier \citep{lih19}. The eight parameters are from or can be derived from typical solar wind observations, such as the magnetic field strength, proton density and temperature, the solar wind speed, and alpha particle density. The four types of solar winds include coronal-hole-origin (CHO) wind, streamer-belt-origin (SBO) wind, sector-reversal-region (SRR) wind, and ejecta.

The CHO wind refers to the fast solar wind coming from the open field lines in a coronal hole, with a speed generally $>$ 500 km s$^{-1}$ at 1 au \citep{she76p71,cra02p29,mcc08p35,cra09p03}. It is characterized typically by a high proton temperature, low plasma density, and outward propagating Alfv\'en waves \citep{sch06p51}. Moreover, this type of wind usually has a relatively steady alpha particle abundance, and the alpha particles often stream faster than protons with a differential velocity comparable to the local Alfv\'en velocity \citep{mar82p35,fuh18p84}. The SBO wind refers to the plasma originating from either the edge of a coronal hole near a streamer belt or the edge of an open streamer, while the SRR wind involves the plasma from the tip of the open streamer where a magnetic sector reversal exists \citep{gos81p38,ant05p99,mar06p01,fou09p89}. Both the SBO and SRR winds contribute to the slow solar wind with a speed often $\lesssim$ 400 km s$^{-1}$ \citep{sch06p51,xuf15p70}. Compared with the CHO wind, they are more variable and filamentary, and have a low proton temperature, high plasma density, low alpha particle abundance, and small alpha$-$proton differential velocity \citep{sch06p51,fuh18p84}. The ejecta concerns the transient wind denoted as coronal mass ejections that may prevail during the solar maximum \citep{sch06p51,che11p01}.

One relevant issue is that the types of solar winds are assessed by probabilities in the categorization scheme. Probabilistic approaches to $k$-Nearest Neighbor classification have already been proposed by many authors \citep[e.g.,][]{hol02p95,tom11p00}. The latest scikit-learn package of python can derive a probability for $k$-Nearest Neighbor classifier, and the probability can be given as
\begin{eqnarray}
P_{ij}=\frac{exp(-d_{ij})}{\sum\!{_{k}}\;{exp(-d_{ik})}},
\end{eqnarray}
where $d_{ij}$ is the Euclidean distance between points $i$ and $j$. The types with probabilities greater than 0.7 are selected to reduce the uncertainties in this paper.

In addition, we discard any observation with ${V}_d/{V}_p < 1\%$ since in the case ${V}_d$ would have a large uncertainty \citep{kas06p05,alt18p12}. Finally, the sample number is about $3.5 \times 10^5$ (25\%) for the CHO wind, $6.5 \times 10^5$(47\%) for the SBO wind, $3.1 \times 10^5$ (22\%) for the SRR wind, $8.2 \times 10^4$ (6\%) for the ejecta during 2005$-$2015. Note that (1) the ejecta has the smallest sample number, and its physical situation is complicated; (2) the SBO wind has the largest sample number, but it seems to be a transition between the CHO wind and the SRR wind based on our primary test; (3) the CHO and SRR winds have the comparable sample numbers. Consequently, only solar winds sorted as CHO and SRR winds will be presented to illustrate the main results in the present paper.

Figure 1 plots the data ordered by the radial angle of ${\bm V}_d$ with a bin of $10^\circ$, where open and filled circles represent the CHO and SRR types, respectively. The radial angle is the angle between ${\bm V}_d$ and ${\bm R}$ (the radial vector of the Sun), defined as
\begin{eqnarray}
\theta_d=\frac{180^\circ}{\pi}\arccos(\frac{{\bm V}_d\cdot{\bm R}}{|{\bm V}_d||{\bm R}|}),
\end{eqnarray}
where $\arccos$ refers to the branch of the inverse cosine function with range [0, $\pi$]; an angle $< 90^\circ$ means ${\bm V}_d$ pointing outward from the Sun while an angle $> 90^\circ$ denotes it toward the Sun. One can see that there are a considerable number of observations with ${\bm V}_d$ directed toward the Sun for the CHO wind, and most observations of the SRR wind exhibit ${\bm V}_d$ toward the Sun.

\begin{figure}
\epsscale{0.9} \plotone{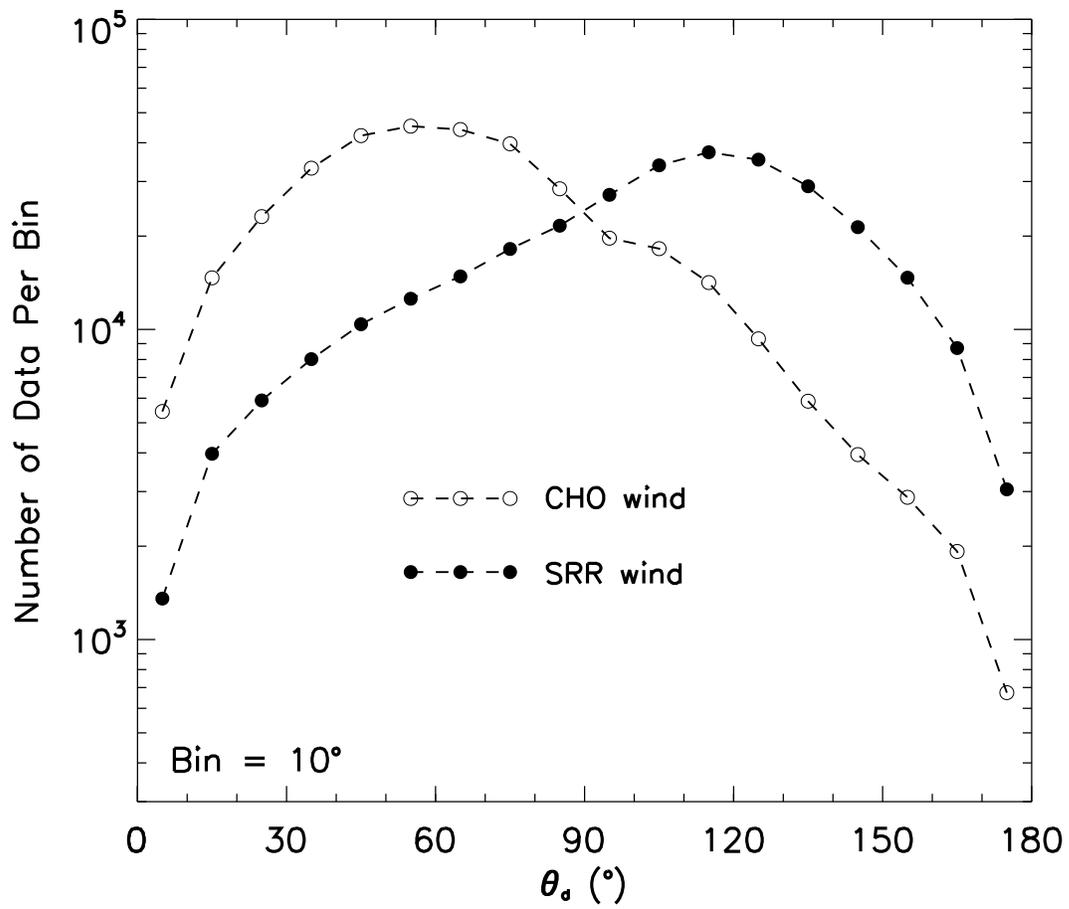} \caption{Data distributions regulated by the radial angle $\theta_d$, where open and filled circles represent the CHO and SRR winds, respectively.}
\end{figure}

The survey of ECWs is carried out by an automatic wave detection procedure that was developed by \citet{zha17p79,zha18p15}. The procedure mainly consists of three steps. The first step is to calculate the reduced magnetic helicity spectrum in the frequency range from 0.05 to 1 Hz for a given magnetic field interval \citep[e.g.,][]{mat82p11,hej11p85}. If the spectrum has absolute values $\geq$ 0.7 in some frequency band (with a minimum bandwidth of 0.05 Hz), the second step will begin with identifying enhanced power spectrum. The enhancement requires transverse wave power three times larger than the background power in the same frequency band. When the above two steps are fulfilled, the third step follows to record the wave with an amplitude criterion of 0.1 nT; the wave amplitude is obtained with employing a band-pass filter technique \citep{wil09p06}. Note that the magnetic helicity is described with respect to the direction of the local background magnetic field in the procedure, and a negative (positive) helicity implies LH (RH) polarization. The magnetic field data are first converted into a field-aligned coordinate system with the $z$ direction along the direction of the background magnetic field (i.e., an average field over the period of the interval) before the magnetic helicity is calculated. This operation removes the inversion of the sign of magnetic helicity when the magnetic field direction changes from Sunward to anti-Sunward or vice versa depending on, for instance, the sector structure of the solar wind.

\section{Results}
The occurrence rate and polarization sense should be two important physical parameters to understand ECWs concerning their generation mechanism \citep{zha19p75}. The wave detection procedure described in Section 2 can give the time intervals of ECWs occurrence, and therefore allows us to calculate their occurrence rate. The polarization sense of ECWs can also be determined directly by the sign of the spectrum values of magnetic helicity. Figure 2 presents the occurrence rates of ECWs (left panel) and the percentages of LH ECWs (right panel) regulated by the radial angle of ${\bm V}_d$, where open and filled circles are for CHO and SRR winds, respectively. The percentages refer to the ratio of the number of time intervals with LH ECWs to the total number of time intervals with either LH or RH ECWs.

Figure 2 shows that the occurrence rates and the percentages of LH ECWs significantly depend on the radial angle $\theta_d$. The occurrence rate for the CHO wind as well as that for the SRR wind decreases from the maximum in the  0$-$10$^\circ$ bin to the minimum at $\theta_d \sim 90^\circ$ and then slightly increases with $\theta_d$. The percentage of LH ECWs with $\theta_d < 90^\circ$ is usually larger than that with $\theta_d > 90^\circ$ for ether the CHO wind or the SRR wind. In both cases the percentages generally exceed $50\%$ when $\theta_d < 90^\circ$, and there are enhancements of the percentages at radial angle around $50^\circ$. As $\theta_d$ increases from about $90^\circ$ to $180^\circ$, the percentage rapidly reduces from about $87.2\%$ to $42.6\%$ for the CHO wind, and it fluctuates around $40\%$ with a minimum of $33.8\%$ for the SRR wind. This means that LH ECWs can become secondary when ${\bm V}_d$ is directed toward the Sun, especially in the SRR wind.

\begin{figure}
\epsscale{0.9} \plotone{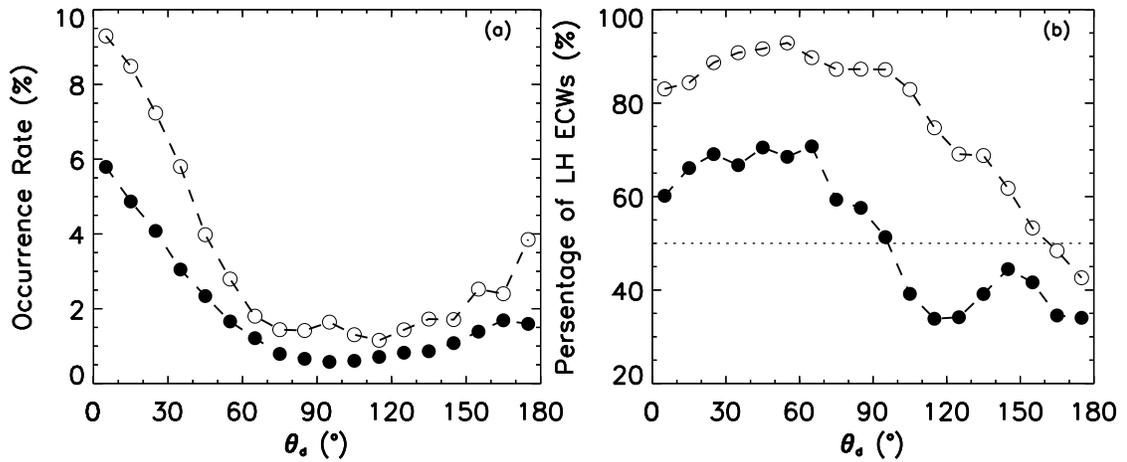} \caption{Occurrence rates of ECWs (left panel) and percentages of LH ECWs  (right panel) with respect to the radial angle $\theta_d$, where open and filled circles represent the CHO and SRR winds, respectively. The dotted line in right panel indicates a value of $50\%$.}
\end{figure}

\begin{figure}
\epsscale{0.9} \plotone{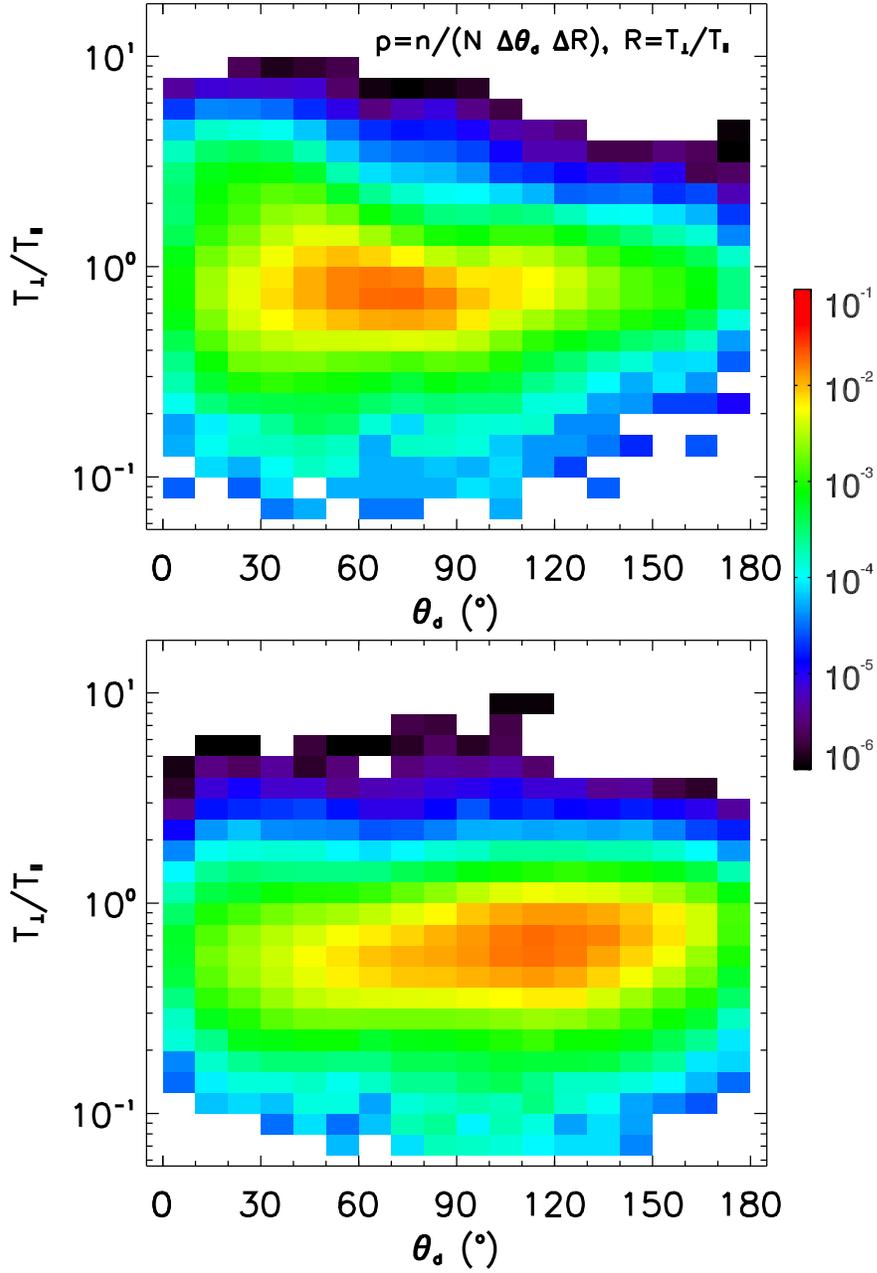} \caption{Probability density distributions $p$($\theta_d$, $T_\perp/T_\parallel$) for the CHO wind (top panel) and the SRR wind (bottom panel), respectively.}
\end{figure}

\begin{figure}
\epsscale{0.9} \plotone{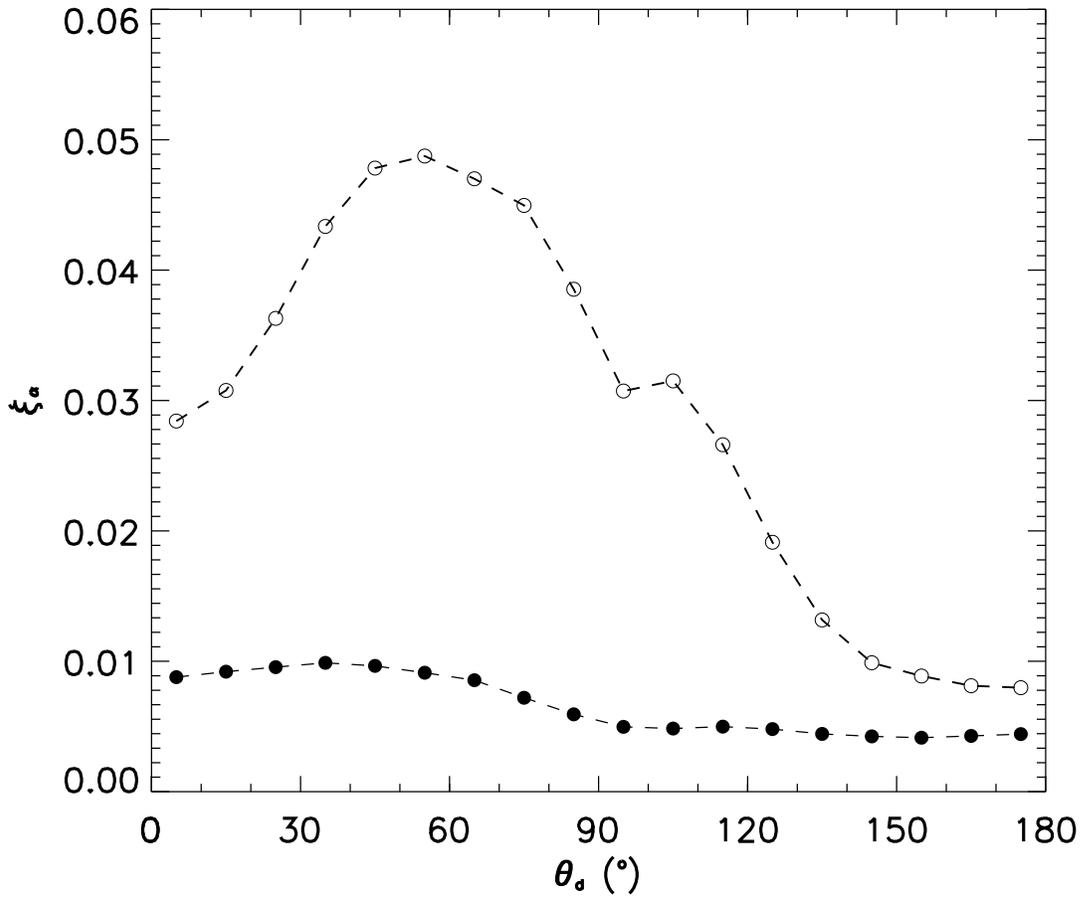} \caption{Medians of kinetic energy $\xi_{\alpha}$ with respect to the radial angle $\theta_d$, where open and filled circles represent the CHO and SRR winds, respectively.}
\end{figure}

\begin{figure}
\epsscale{0.9} \plotone{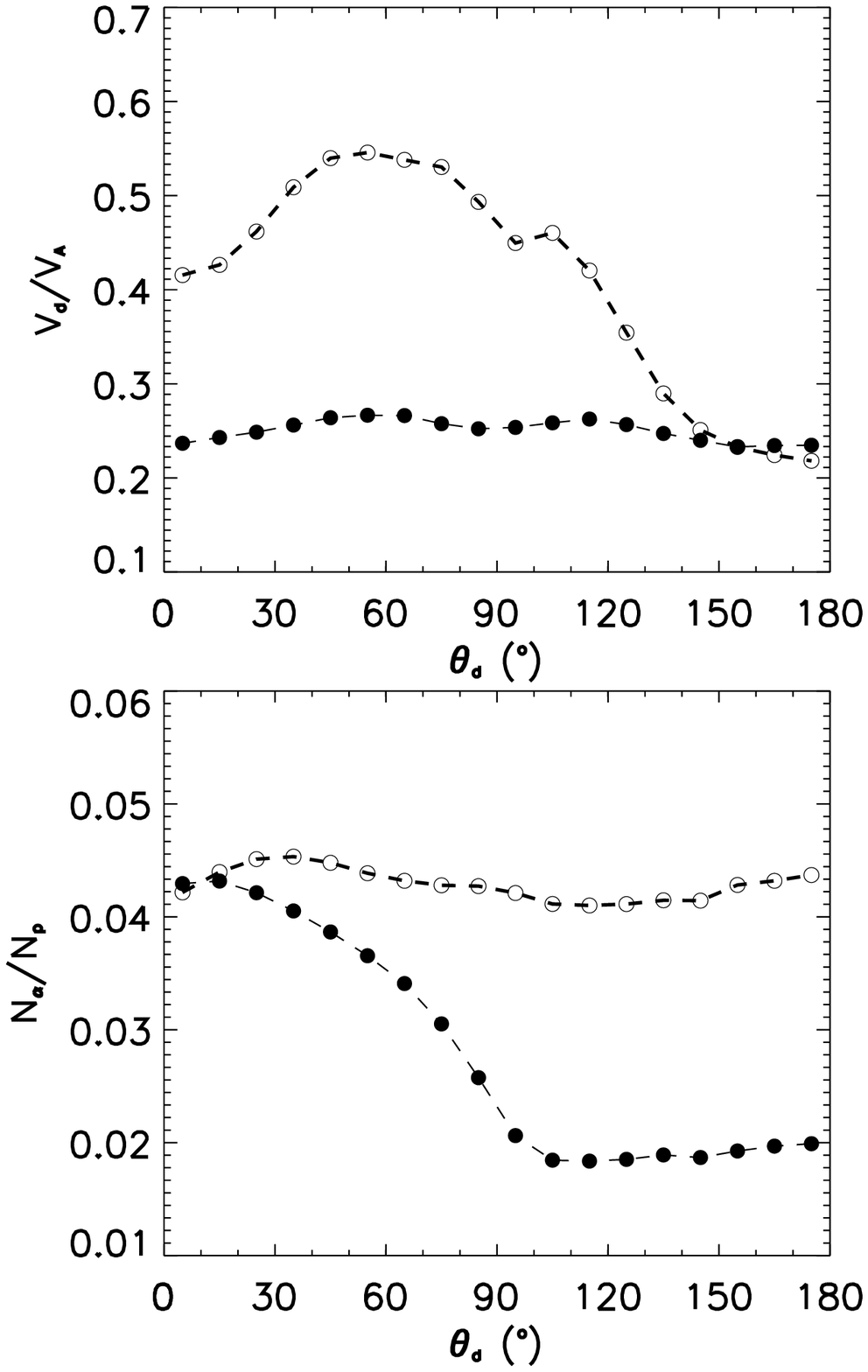} \caption{Medians of $V_d/V_A$ (top panel) and $N_{\alpha}/N_{p}$ (bottom panel) with respect to the radial angle $\theta_d$, where open and filled circles represent the CHO and SRR winds, respectively.}
\end{figure}

In order to understand the implication of results presented in Figure 2, proton temperature anisotropies and parameters for alpha particles are investigated. The temperature anisotropies are described by $T_\perp/T_\parallel$, where $T_{\perp}$ and $T_\parallel$ are proton temperatures perpendicular and parallel to the background magnetic field, respectively. Figure 3 plots the probability density distributions $p$($\theta_d$, $T_\perp/T_\parallel$) for the CHO wind (top panel) and the SRR wind (bottom panel), respectively. Here the expression $p=n/(N\Delta\theta_d\Delta{R})$ is used, where $n$ and $N$ are the sample number in each cell and the total sample number in each panel, respectively, $\Delta\theta_d\Delta{R}$ represents the size of the cell with $R=T_\perp/T_\parallel$. One may first find that proton temperature anisotropies are common in solar wind plasmas. This means that ECWs will be excited by the temperature-anisotropy-driven instabilities once their threshold conditions are fulfilled. Note that the distributions of $T_\perp/T_\parallel$ can be different for different $\theta_d $ as well as for different types of solar winds. The plasma of the CHO wind with $\theta_d < 90^{\circ}$ is characterized by the widest distribution of $T_\perp/T_\parallel$ relative to other plasmas.

The parameters for alpha particles investigated in this study mainly include the density and differential velocity. It has been found that the kinetic energy ratio, defined by
\begin{eqnarray}
\xi_{\alpha} = \frac{m_{\alpha}N_{\alpha}V_d^2}{m_pN_{p}V_A^2},
\end{eqnarray}
is a relevant parameter to discuss the occurrence of ECWs \citep{zha19p75}, where $m_{\alpha}$ and $N_{\alpha}$ ($m_{p}$ and $N_{p}$) are the mass and number density of alpha particles (protons), $V_A$ is the local Alfv\'en velocity. Figure 4 displays medians of $\xi_{\alpha}$ with respect to $\theta_d$. Several points can be found as follows. Firstly, the median of $\xi_{\alpha}$ for the CHO wind is always larger than that for the SRR wind in a given radial angle bin. Secondly, the median of $\xi_{\alpha}$ in the 0$-$10$^\circ$ bin is larger than that in the 170$-$180$^\circ$ bin regardless the solar wind types. Thirdly, there is a distinct maximum with $\theta_d \sim 50^{\circ}$ for the CHO wind.

The parameter $\xi_{\alpha}$ may represent the kinetic energy of alpha particle flow in the proton reference frame; it is normalized by the kinetic energy of protons with a bulk velocity $V_A$ for the sake of convenience. Note that usually only the parameter $V_d/V_A$ is investigated in existing literatures in which a fixed $N_{\alpha}/N_{p}$ is used \citep[e.g.,][]{hel06p07,pod11p41}. The proposal of $\xi_{\alpha}$ in the present paper should be clarified. For this purpose Figure 5 is plotted, where top panel is for $V_d/V_A$ and bottom panel is for $N_{\alpha}/N_{p}$ with respect to the radial angle $\theta_d$. It is found that the wave occurrence rates shown in Figure 2 cannot be well understood just in terms of $V_d/V_A$ (or just by $N_{\alpha}/N_{p}$). The median of $V_d/V_A$ for SRR wind (filled circles in top panel) is approximately a constant around 0.24, and it is nearly equal to or even exceeds that for CHO wind when $\theta_d > 140^\circ$. Here one may expect that an approximately constant wave occurrence rate should arise irrespective of $\theta_d$ for the SRR wind, and comparable occurrence rates between the CHO and SRR winds would happen if $\theta_d > 140^\circ$. However, it is not the case since the occurrence rate for SRR wind is considerably higher when $\theta_d$ is small, and on the other hand it is much lower than that for CHO wind when $\theta_d$ approaches $180^\circ$. This disagreement seems to be removed by the changing $N_{\alpha}/N_{p}$ for SRR wind shown in bottom panel of Figure 5, because the median of $N_{\alpha}/N_{p}$ for SRR wind is greatly higher when $\theta_d$ is small, and it is significantly lower than that for CHO wind when $\theta_d$ is large. Moreover, it is revealed that large occurrence rates take place mainly at regions with larger $V_d/V_A$ and/or higher $N_{\alpha}/N_{p}$ when occurrence rates of ECWs are investigated in the space of $(V_d/V_A,N_{\alpha}/N_{p})$ \citep{zha19p75}. These results seem to imply that both $V_d/V_A$ and $N_{\alpha}/N_{p}$ are simultaneously relevant to the occurrence of the ECWs, which promotes us to speculate that an integrated parameter consisting of $V_d/V_A$ and $N_{\alpha}/N_{p}$ should be appropriate. Consequently the kinetic energy ratio $\xi_{\alpha}$ is proposed to discuss the occurrence of the ECWs. One should keep in mind that this proposal is based on our analyses of observation data since a specific theory concerning $\xi_{\alpha}$ is absent.

\section{Discussion and Conclusion}
Two points revealed by Figure 2(a) should be notable. One is that the occurrence rate for the CHO wind is always higher than that for the SRR wind at a fixed radial angle bin. The other is that the occurrence rate in the 0$-$10$^\circ$ bin is much higher than that in the 170$-$180$^\circ$ bin for either the CHO wind or the SRR wind. We propose here that the temperature-anisotropy-driven instabilities with effect of alpha$-$proton differential flow are likely responsible for the generation of ECWs in the solar wind.
The presence of alpha$-$proton differential flow contributes to the excitation of the instabilities; a faster differential flow gives rise to a more rapid growth of the instabilities \citep{pod11p41}, and therefore higher occurrence rates of ECWs in observations. Based on this conception, both points in Figure 2(a) can be understood since Figure 4 shows that (1) statistically the kinetic energy (represented by an energy ratio defined by Equation (3)) of the differential flow for the CHO wind is always larger than that for the SRR wind in a given radial angle bin; (2) the kinetic energy in the 0$-$10$^\circ$ bin is also greater than that in the 170$-$180$^\circ$ bin for either the CHO wind or the SRR wind.

In the above discussion one may realize that there is still an inconsistence between occurrence rates and kinetic energies when the radial angle approaches $90^\circ$. For a given solar wind type the occurrence rate in this region is much low relative to that in other regions while it is not the case for the kinetic energy. We speculate this is due to the limit from observation. A lot of ECWs would possibly not be recognized when the radial angle is around $90^\circ$. In this region the spacecraft probably crosses approximately perpendicularly to the wave vector of ECWs and fails to detect the variation of the wave fields. Here we refer to that the differential flow is believed to be aligned with the ambient magnetic field \citep{kas06p05,fuh18p84}, and meanwhile the ECWs usually propagate nearly parallel to the magnetic field \citep{jia09p05,zha18p15}.

The result concerning percentages of LH ECWs in this paper may provide crucial indication for the effect of alpha$-$proton differential flow on propagation direction of ECWs. Figure 2(b) shows that the percentages of LH ECWs are significantly different between the case of differential flow with direction outward from the Sun and that with direction toward the Sun, which is consistent with the theory by \citet{pod11p41}. According to the theory, the alpha$-$proton differential flow with direction outward from the Sun causes proton cyclotron instability ($T_{\perp} > T_\parallel$) to preferentially generate cyclotron waves propagating away from the Sun, and it leads parallel firehose instability ($T_{\perp} < T_\parallel$) to preferentially generate magnetosonic waves propagating toward the Sun. Note that magnetosonic waves are inherently RH waves in the plasma frame, but these waves shall appear as LH waves in the spacecraft frame because their polarization will be reversed in this reference frame by large doppler shifts due to the fast motion of the solar wind \citep{jia09p05,gar16p30}. On the other hand, cyclotron waves excited by the proton cyclotron instability tend to propagate toward the Sun while magnetosonic waves produced by the firehose instability will propagate preferentially outward once the differential flow is toward the Sun.
Consequently, ECWs generated by these two instabilities will be dominated by LH polarization when the differential flow is directed outward, and they will favor RH polarization when the direction of differential flow is toward the Sun.

In addition, there are considerable enhancements of the percentages at radial angle around $50^\circ$ (Figure 2(b)), and meanwhile the kinetic energies of the differential flow are larger at the same radial angle. This coincidence should be meaningful and can reinforce the indication for the effect of alpha$-$proton differential flow since the larger kinetic energies of the differential flow at radial angle around $50^\circ$ would result in more LH ECWs and therefore larger percentages of LH ECWs. Here, one may also note that the percentage is still higher than $50\%$  even though the radial angle is greater than $90^\circ$ for the CHO wind. This phenomenon, unfortunately, can not be understood directly in the present paper.
A reason might be the presence of proton$-$proton differential flow in the CHO wind, which could make the polarization complicated in intuition. One notable result is that the percentage of LH ECWs rapidly decreases as the radial angle approaches $180^\circ$ for the CHO wind. An investigation on the role of proton$-$proton differential flow is beyond the scope of this paper and is desirable for future study.

Observations of LH ECWs with a percentage less than $50\%$ in the solar wind are rarely reported in previous literatures; to the best of our knowledge, only \citet{zha18p15} reported the percentages less than $50\%$ in 6 months among 84 months based on the \emph{STEREO} mission. This should be because previous surveys did not discriminate the directions of alpha$-$proton differential flows. In the case the dominance of RH ECWs would be hidden by more LH ECWs from the region with the outward directed differential flows. The result in Figure 2(b) reveals that a low percentage of LH ECWs can arise for the CHO wind when the radial angle is near $180^\circ$, and it is particularly common for the SRR wind once the radial angle exceeds $90^\circ$. In this regard, this paper tend to present a condition for RH ECWs dominating. That is a plasma with the alpha$-$proton differential flow directed toward the Sun.

In conclusion, this paper reveals that the occurrence rates and polarization senses of ECWs significantly depend on the direction of alpha$-$proton differential flow in the solar wind. It is shown that the dominant polarization is LH polarization when the differential flow points outward from the Sun, while it can be RH polarization when the differential flow is with direction toward the Sun. Further investigation on proton and alpha particles illustrates that large kinetic energies of the differential flow correspond to high occurrence rates of ECWs. These results are well in line with the theory for effects of alpha$-$proton differential flow on proton temperature anisotropy instabilities.

\acknowledgments
The authors thank the SWE team and MFI team on \emph{Wind} for providing the data, which are available via the Coordinated Data Analysis Web (http://cdaweb.gsfc.nasa.gov/cdaweb/istp$_-$public/). This research was supported by NSFC under grant Nos. 41874204, 41874203, 41574169, 41674170, 41531071, 41804163. Research by G. Q. Zhao was also supported by the Project for Scientific Innovation Talent in Universities of Henan Province (19HASTIT020). Research by H. Li was also supported by Young Elite Scientists Sponsorship Program by CAST, Youth Innovation Promotion Association of the Chinese Academy of Sciences, and in part by the Specialized Research Fund for State Key Laboratories of China.


\begin{thebibliography}{69}
\expandafter\ifx\csname natexlab\endcsname\relax\def\natexlab#1{#1}\fi

\bibitem[{{Alterman} {et~al.}(2018){Alterman}, {Kasper}, {Stevens}, \&
  {Koval}}]{alt18p12}
{Alterman}, B.~L., {Kasper}, J.~C., {Stevens}, M.~L., \& {Koval}, A. 2018,
  \apj, 864, 112

\bibitem[{{Antonucci} {et~al.}(2005){Antonucci}, {Abbo}, \&
  {Dodero}}]{ant05p99}
{Antonucci}, E., {Abbo}, L., \& {Dodero}, M.~A. 2005, \aap, 435, 699

\bibitem[{{Asbridge} {et~al.}(1976){Asbridge}, {Bame}, {Feldman}, \&
  {Montgomery}}]{asb76p19}
{Asbridge}, J.~R., {Bame}, S.~J., {Feldman}, W.~C., \& {Montgomery}, M.~D.
  1976, \jgr, 81, 2719

\bibitem[{{Boardsen} {et~al.}(2015){Boardsen}, {Jian}, {Raines}, {Gershman},
  {Zurbuchen}, {Roberts}, \& {Korth}}]{boa15p10}
{Boardsen}, S.~A., {Jian}, L.~K., {Raines}, J.~L., {et~al.} 2015, Journal of
  Geophysical Research (Space Physics), 120, 10207

\bibitem[{{Chen}(2011)}]{che11p01}
{Chen}, P.~F. 2011, Living Reviews in Solar Physics, 8, 1

\bibitem[{{Cranmer}(2002)}]{cra02p29}
{Cranmer}, S.~R. 2002, \ssr, 101, 229

\bibitem[{{Cranmer}(2009)}]{cra09p03}
---. 2009, Living Reviews in Solar Physics, 6, 3

\bibitem[{{D'Amicis} {et~al.}(2019){D'Amicis}, {Matteini}, \&
  {Bruno}}]{ami19p65}
{D'Amicis}, R., {Matteini}, L., \& {Bruno}, R. 2019, \mnras, 483, 4665

\bibitem[{{Foullon} {et~al.}(2009){Foullon}, {Lavraud}, {Wardle}, {Owen},
  {Kucharek}, {Fazakerley}, {Larson}, {Lucek}, {Luhmann}, {Opitz}, {Sauvaud},
  \& {Skoug}}]{fou09p89}
{Foullon}, C., {Lavraud}, B., {Wardle}, N.~C., {et~al.} 2009, \solphys, 259,
  389

\bibitem[{{Fu} {et~al.}(2018){Fu}, {Madjarska}, {Li}, {Xia}, \&
  {Huang}}]{fuh18p84}
{Fu}, H., {Madjarska}, M.~S., {Li}, B., {Xia}, L., \& {Huang}, Z. 2018, \mnras,
  478, 1884

\bibitem[{{Gary}(1993)}]{gar93}
{Gary}, S.~P. 1993, {Theory of Space Plasma Microinstabilities}, 193

\bibitem[{Gary(2015)}]{Gar15p49}
Gary, S.~P. 2015, Philosophical Transactions of the Royal Society of London A:
  Mathematical, Physical and Engineering Sciences, 373, 40149

\bibitem[{{Gary} {et~al.}(2016){Gary}, {Jian}, {Broiles}, {Stevens}, {Podesta},
  \& {Kasper}}]{gar16p30}
{Gary}, S.~P., {Jian}, L.~K., {Broiles}, T.~W., {et~al.} 2016, Journal of
  Geophysical Research (Space Physics), 121, 30

\bibitem[{{Gary} {et~al.}(1976){Gary}, {Montgomery}, {Feldman}, \&
  {Forslund}}]{gar76p41}
{Gary}, S.~P., {Montgomery}, M.~D., {Feldman}, W.~C., \& {Forslund}, D.~W.
  1976, \jgr, 81, 1241

\bibitem[{{Gary} {et~al.}(2000{\natexlab{a}}){Gary}, {Yin}, {Winske}, \&
  {Reisenfeld}}]{gar00p20}
{Gary}, S.~P., {Yin}, L., {Winske}, D., \& {Reisenfeld}, D.~B.
  2000{\natexlab{a}}, \jgr, 105, 20

\bibitem[{{Gary} {et~al.}(2000{\natexlab{b}}){Gary}, {Yin}, {Winske}, \&
  {Reisenfeld}}]{gar00p55}
---. 2000{\natexlab{b}}, \grl, 27, 1355

\bibitem[{{Gosling} {et~al.}(1981){Gosling}, {Borrini}, {Asbridge}, {Bame},
  {Feldman}, \& {Hansen}}]{gos81p38}
{Gosling}, J.~T., {Borrini}, G., {Asbridge}, J.~R., {et~al.} 1981, \jgr, 86,
  5438

\bibitem[{{He} {et~al.}(2011){He}, {Marsch}, {Tu}, {Yao}, \& {Tian}}]{hej11p85}
{He}, J., {Marsch}, E., {Tu}, C., {Yao}, S., \& {Tian}, H. 2011, \apj, 731, 85

\bibitem[{{He} {et~al.}(2015){He}, {Wang}, {Tu}, {Marsch}, \&
  {Zong}}]{hej15p31}
{He}, J., {Wang}, L., {Tu}, C., {Marsch}, E., \& {Zong}, Q. 2015, \apjl, 800,
  L31

\bibitem[{{He} {et~al.}(2018){He}, {Zhu}, {Chen}, {Salem}, {Stevens}, {Li},
  {Ruan}, {Zhang}, \& {Tu}}]{hej18p48}
{He}, J., {Zhu}, X., {Chen}, Y., {et~al.} 2018, \apj, 856, 148

\bibitem[{{Hellinger} \& {Tr{\'a}vn{\'{\i}}{\v c}ek}(2006)}]{hel06p07}
{Hellinger}, P., \& {Tr{\'a}vn{\'{\i}}{\v c}ek}, P. 2006, Journal of
  Geophysical Research (Space Physics), 111, A01107

\bibitem[{{Hellinger} {et~al.}(2006){Hellinger}, {Tr{\'a}vn{\'{\i}}{\v c}ek},
  {Kasper}, \& {Lazarus}}]{hel06p01}
{Hellinger}, P., {Tr{\'a}vn{\'{\i}}{\v c}ek}, P., {Kasper}, J.~C., \&
  {Lazarus}, A.~J. 2006, \grl, 33, L09101

\bibitem[{{Hollweg}(1975)}]{hol75p63}
{Hollweg}, J.~V. 1975, Reviews of Geophysics and Space Physics, 13, 263

\bibitem[{{Holmes} \& {Adams}(2002)}]{hol02p95}
{Holmes}, C.~C., \& {Adams}, N.~M. 2002, Journal of the Royal Statistical
  Society: Series B (Statistical Methodology), 64, 295

\bibitem[{{Hu} \& {Rifai Habbal}(1999)}]{huy99p45}
{Hu}, Y.~Q., \& {Rifai Habbal}, S. 1999, \jgr, 104, 17045

\bibitem[{{Jian} {et~al.}(2010){Jian}, {Russell}, {Luhmann}, {Anderson},
  {Boardsen}, {Strangeway}, {Cowee}, \& {Wennmacher}}]{jia10p15}
{Jian}, L.~K., {Russell}, C.~T., {Luhmann}, J.~G., {et~al.} 2010, Journal of
  Geophysical Research (Space Physics), 115, A12115

\bibitem[{{Jian} {et~al.}(2009){Jian}, {Russell}, {Luhmann}, {Strangeway},
  {Leisner}, \& {Galvin}}]{jia09p05}
---. 2009, \apjl, 701, L105

\bibitem[{{Jian} {et~al.}(2014){Jian}, {Wei}, {Russell}, {Luhmann}, {Klecker},
  {Omidi}, {Isenberg}, {Goldstein}, {Figueroa-Vi{\~n}as}, \&
  {Blanco-Cano}}]{jia14p23}
{Jian}, L.~K., {Wei}, H.~Y., {Russell}, C.~T., {et~al.} 2014, \apj, 786, 123

\bibitem[{{Kasper} {et~al.}(2002){Kasper}, {Lazarus}, \& {Gary}}]{kas02p39}
{Kasper}, J.~C., {Lazarus}, A.~J., \& {Gary}, S.~P. 2002, \grl, 29, 1839

\bibitem[{{Kasper} {et~al.}(2006){Kasper}, {Lazarus}, {Steinberg}, {Ogilvie},
  \& {Szabo}}]{kas06p05}
{Kasper}, J.~C., {Lazarus}, A.~J., {Steinberg}, J.~T., {Ogilvie}, K.~W., \&
  {Szabo}, A. 2006, Journal of Geophysical Research (Space Physics), 111,
  A03105

\bibitem[{{Lepping} {et~al.}(1995){Lepping}, {Ac{\~u}na}, {Burlaga}, {Farrell},
  {Slavin}, {Schatten}, {Mariani}, {Ness}, {Neubauer}, {Whang}, {Byrnes},
  {Kennon}, {Panetta}, {Scheifele}, \& {Worley}}]{lep95p07}
{Lepping}, R.~P., {Ac{\~u}na}, M.~H., {Burlaga}, L.~F., {et~al.} 1995, \ssr,
  71, 207

\bibitem[{{Li} {et~al.}(2019{\natexlab{a}}){Li}, {Wang}, {Tu}, \& {Xu}}]{lih19}
{Li}, H., {Wang}, C., {Tu}, C., \& {Xu}, F. 2019{\natexlab{a}}, Earth and Space
  Sciecne, submitted

\bibitem[{{Li} {et~al.}(2019{\natexlab{b}}){Li}, {Yang}, {Wu}, \&
  {Wang}}]{liq19p55}
{Li}, Q.~H., {Yang}, L., {Wu}, D.~J., \& {Wang}, T.~Y. 2019{\natexlab{b}},
  \apj, 874, 55

\bibitem[{{Li} \& {Habbal}(2000)}]{lix00p83}
{Li}, X., \& {Habbal}, S.~R. 2000, \jgr, 105, 7483

\bibitem[{{Lu} {et~al.}(2006){Lu}, {Xia}, \& {Wang}}]{luq06p01}
{Lu}, Q.~M., {Xia}, L.~D., \& {Wang}, S. 2006, Journal of Geophysical Research
  (Space Physics), 111, A09101

\bibitem[{{Markovskii} {et~al.}(2018){Markovskii}, {Chandran}, \&
  {Vasquez}}]{mar18p53}
{Markovskii}, S.~A., {Chandran}, B.~D.~G., \& {Vasquez}, B.~J. 2018, \apj, 856,
  153

\bibitem[{{Marsch}(2006)}]{mar06p01}
{Marsch}, E. 2006, Living Reviews in Solar Physics, 3, 1

\bibitem[{{Marsch} {et~al.}(2004){Marsch}, {Ao}, \& {Tu}}]{mar04p02}
{Marsch}, E., {Ao}, X.-Z., \& {Tu}, C.-Y. 2004, Journal of Geophysical Research
  (Space Physics), 109, A04102

\bibitem[{{Marsch} {et~al.}(1982{\natexlab{a}}){Marsch}, {Goertz}, \&
  {Richter}}]{mar82p30}
{Marsch}, E., {Goertz}, C.~K., \& {Richter}, K. 1982{\natexlab{a}}, \jgr, 87,
  5030

\bibitem[{{Marsch} {et~al.}(1981){Marsch}, {Rosenbauer}, {Schwenn},
  {Muehlhaeuser}, \& {Denskat}}]{mar81p99}
{Marsch}, E., {Rosenbauer}, H., {Schwenn}, R., {Muehlhaeuser}, K.-H., \&
  {Denskat}, K.~U. 1981, \jgr, 86, 9199

\bibitem[{{Marsch} {et~al.}(1982{\natexlab{b}}){Marsch}, {Rosenbauer},
  {Schwenn}, {Muehlhaeuser}, \& {Neubauer}}]{mar82p35}
{Marsch}, E., {Rosenbauer}, H., {Schwenn}, R., {Muehlhaeuser}, K.-H., \&
  {Neubauer}, F.~M. 1982{\natexlab{b}}, \jgr, 87, 35

\bibitem[{{Matthaeus} \& {Goldstein}(1982)}]{mat82p11}
{Matthaeus}, W.~H., \& {Goldstein}, M.~L. 1982, \jgr, 87, 6011

\bibitem[{{McComas} {et~al.}(2008){McComas}, {Ebert}, {Elliott}, {Goldstein},
  {Gosling}, {Schwadron}, \& {Skoug}}]{mcc08p35}
{McComas}, D.~J., {Ebert}, R.~W., {Elliott}, H.~A., {et~al.} 2008, \grl, 35,
  L18103

\bibitem[{{Ogilvie} {et~al.}(1995){Ogilvie}, {Chornay}, {Fritzenreiter},
  {Hunsaker}, {Keller}, {Lobell}, {Miller}, {Scudder}, {Sittler}, {Torbert},
  {Bodet}, {Needell}, {Lazarus}, {Steinberg}, {Tappan}, {Mavretic}, \&
  {Gergin}}]{ogi95p55}
{Ogilvie}, K.~W., {Chornay}, D.~J., {Fritzenreiter}, R.~J., {et~al.} 1995,
  \ssr, 71, 55

\bibitem[{{Omidi} {et~al.}(2014){Omidi}, {Isenberg}, {Russell}, {Jian}, \&
  {Wei}}]{omi14p42}
{Omidi}, N., {Isenberg}, P., {Russell}, C.~T., {Jian}, L.~K., \& {Wei}, H.~Y.
  2014, Journal of Geophysical Research (Space Physics), 119, 1442

\bibitem[{{Podesta} \& {Gary}(2011)}]{pod11p41}
{Podesta}, J.~J., \& {Gary}, S.~P. 2011, \apj, 742, 41

\bibitem[{{Robbins} {et~al.}(1970){Robbins}, {Hundhausen}, \&
  {Bame}}]{rob70p78}
{Robbins}, D.~E., {Hundhausen}, A.~J., \& {Bame}, S.~J. 1970, \jgr, 75, 1178

\bibitem[{{Rodr{\'{\i}}guez-Mart{\'{\i}}nez}
  {et~al.}(2010){Rodr{\'{\i}}guez-Mart{\'{\i}}nez}, {Blanco-Cano}, {Russell},
  {Leisner}, {Wilson}, \& {Dougherty}}]{rod10p07}
{Rodr{\'{\i}}guez-Mart{\'{\i}}nez}, M., {Blanco-Cano}, X., {Russell}, C.~T.,
  {et~al.} 2010, Journal of Geophysical Research (Space Physics), 115, A09207

\bibitem[{{Russell} \& {Blancocano}(2007)}]{rus07p23}
{Russell}, C., \& {Blancocano}, X. 2007, Journal of Atmospheric and
  Solar-Terrestrial Physics, 69, 1723

\bibitem[{{Schwartz} {et~al.}(1996){Schwartz}, {Burgess}, \&
  {Moses}}]{sch96p34}
{Schwartz}, S.~J., {Burgess}, D., \& {Moses}, J.~J. 1996, Annales Geophysicae,
  14, 1134

\bibitem[{{Schwenn}(2006)}]{sch06p51}
{Schwenn}, R. 2006, \ssr, 124, 51

\bibitem[{{Sheeley} {et~al.}(1976){Sheeley}, {Harvey}, \& {Feldman}}]{she76p71}
{Sheeley}, Jr., N.~R., {Harvey}, J.~W., \& {Feldman}, W.~C. 1976, \solphys, 49,
  271

\bibitem[{{Smith} {et~al.}(1985){Smith}, {Goldstein}, {Gary}, \&
  {Russell}}]{smi85p29}
{Smith}, C.~W., {Goldstein}, M.~L., {Gary}, S.~P., \& {Russell}, C.~T. 1985,
  \jgr, 90, 1429

\bibitem[{{Soucek} {et~al.}(2015){Soucek}, {Escoubet}, \& {Grison}}]{sou15p38}
{Soucek}, J., {Escoubet}, C.~P., \& {Grison}, B. 2015, Journal of Geophysical
  Research (Space Physics), 120, 2838

\bibitem[{{Stakhiv} {et~al.}(2015){Stakhiv}, {Landi}, {Lepri}, {Oran}, \&
  {Zurbuchen}}]{sta15p00}
{Stakhiv}, M., {Landi}, E., {Lepri}, S.~T., {Oran}, R., \& {Zurbuchen}, T.~H.
  2015, \apj, 801, 100

\bibitem[{{Stansby} {et~al.}(2019){Stansby}, {Horbury}, \&
  {Matteini}}]{sta19p06}
{Stansby}, D., {Horbury}, T.~S., \& {Matteini}, L. 2019, \mnras, 482, 1706

\bibitem[{{Tomasev} {et~al.}(2011){Tomasev}, {Radovanovic}, \&
  {Mladenic}}]{tom11p00}
{Tomasev}, N., {Radovanovic}, M., \& {Mladenic}, D. 2011, Proceedings of the
  20th ACM Conference on Information and Knowledge Management, CIKM'11, DOI:
  10.1145/2063576.2063919

\bibitem[{{Verscharen} {et~al.}(2013){Verscharen}, {Bourouaine}, \&
  {Chandran}}]{ver13p63}
{Verscharen}, D., {Bourouaine}, S., \& {Chandran}, B.~D.~G. 2013, \apj, 773,
  163

\bibitem[{{Wilson} {et~al.}(2009){Wilson}, {Cattell}, {Kellogg}, {Goetz},
  {Kersten}, {Kasper}, {Szabo}, \& {Meziane}}]{wil09p06}
{Wilson}, III, L.~B., {Cattell}, C.~A., {Kellogg}, P.~J., {et~al.} 2009,
  Journal of Geophysical Research (Space Physics), 114, A10106

\bibitem[{{Wong} {et~al.}(1991){Wong}, {Goldstein}, \& {Smith}}]{won91p85}
{Wong}, H.~K., {Goldstein}, M.~L., \& {Smith}, C.~W. 1991, \jgr, 96, 285

\bibitem[{{Woodham} {et~al.}(2018){Woodham}, {Wicks}, {Verscharen}, \&
  {Owen}}]{woo18p49}
{Woodham}, L.~D., {Wicks}, R.~T., {Verscharen}, D., \& {Owen}, C.~J. 2018,
  \apj, 856, 49

\bibitem[{{Xiang} {et~al.}(2018{\natexlab{a}}){Xiang}, {Wu}, \&
  {Chen}}]{xia18p64}
{Xiang}, L., {Wu}, D.~J., \& {Chen}, L. 2018{\natexlab{a}}, \apj, 869, 64

\bibitem[{{Xiang} {et~al.}(2018{\natexlab{b}}){Xiang}, {Wu}, \&
  {Chen}}]{xia18p08}
---. 2018{\natexlab{b}}, \apj, 857, 108

\bibitem[{{Xu} \& {Borovsky}(2015)}]{xuf15p70}
{Xu}, F., \& {Borovsky}, J.~E. 2015, Journal of Geophysical Research (Space
  Physics), 120, 70

\bibitem[{{Yoon}(2017)}]{yoo17p04}
{Yoon}, P.~H. 2017, Reviews of Modern Plasma Physics, 1, 4

\bibitem[{{Zhao} {et~al.}(2017{\natexlab{a}}){Zhao}, {Chu}, {Lin}, {Yang},
  {Feng}, {Wu}, \& {Liu}}]{zha17p79}
{Zhao}, G.~Q., {Chu}, Y.~H., {Lin}, P.~H., {et~al.} 2017{\natexlab{a}}, Journal
  of Geophysical Research (Space Physics), 122, 4879

\bibitem[{{Zhao} {et~al.}(2017{\natexlab{b}}){Zhao}, {Feng}, {Wu}, {Chu}, \&
  {Huang}}]{zha17p08}
{Zhao}, G.~Q., {Feng}, H.~Q., {Wu}, D.~J., {Chu}, Y.~H., \& {Huang}, J.
  2017{\natexlab{b}}, \apjl, 847, L8

\bibitem[{{Zhao} {et~al.}(2018){Zhao}, {Feng}, {Wu}, {Liu}, {Zhao}, {Zhao}, \&
  {Huang}}]{zha18p15}
{Zhao}, G.~Q., {Feng}, H.~Q., {Wu}, D.~J., {et~al.} 2018, Journal of
  Geophysical Research (Space Physics), 123, 1715

\bibitem[{{Zhao} {et~al.}(2019){Zhao}, {Feng}, {Wu}, {Pi}, \&
  {Huang}}]{zha19p75}
{Zhao}, G.~Q., {Feng}, H.~Q., {Wu}, D.~J., {Pi}, G., \& {Huang}, J. 2019, \apj,
  871, 175

\end{thebibliography}

\end{document}